\documentclass[aps,prb,twocolumn,showpacs,superscriptaddress,preprintnumbers]{revtex4}
\usepackage{graphicx} 
\usepackage{amsmath}
\usepackage{graphicx,epstopdf}

\newcommand{\be}{\begin{equation}}
\newcommand{\ee}{\end{equation}}
\newcommand{\bea}{\begin{eqnarray}}
\newcommand{\eea}{\end{eqnarray}}
\newcommand{\bse}{\begin{subequations}}
\newcommand{\ese}{\end{subequations}}

\begin{document}

\title{Alternating-spin-chain compound AgVOAsO$_{4}$ probed by $^{75}$As NMR}

\author{N. Ahmed}
\affiliation{School of Physics, Indian Institute of Science
Education and Research Thiruvananthapuram-695016, India}
\author{P. Khuntia}
\author{H. Rosner}
\author{M. Baenitz}
\affiliation{Max Planck Institute for Chemical Physics of Solids, 01187 Dresden, Germany}
\author{A. A. Tsirlin}
\affiliation{Max Planck Institute for Chemical Physics of Solids, 01187 Dresden, Germany}
\affiliation{Experimental Physics VI, Center for Electronic Correlations and Magnetism, Institute of Physics, University of Augsburg, 86135 Augsburg, Germany}
\author{R. Nath}
\email{rnath@iisertvm.ac.in}
\affiliation{School of Physics, Indian Institute of Science
Education and Research Thiruvananthapuram-695016, India}
\date{\today}

\begin{abstract}
$^{75}$As NMR measurements were performed on a polycrystalline sample of spin-$\frac{1}{2}$ alternating-spin-chain Heisenberg antiferromagnet AgVOAsO$_{4}$. Temperature-dependent NMR shift $K(T)$, which is a direct measure of the intrinsic spin susceptibility, agrees very well with the spin-$\frac{1}{2}$ alternating-chain model, justifying the assignment of the spin lattice. From the analysis of $K(T)$, magnetic exchange parameters were estimated as follows: the leading exchange $J/k_{\rm B} \simeq 38.4$~K, alternation ratio $\alpha = J'/J \simeq 0.68$, and spin gap $\Delta/k_{\rm B} \simeq 15$~K. The transferred hyperfine coupling between the $^{75}$As nucleus and V$^{4+}$ spins obtained by comparing the NMR shift with bulk susceptibility amounts to $A_{\rm hf} \simeq 3.3$~T/$\mu_{\rm B}$. Our temperature-dependent spin-lattice relaxation rate $1/T_{1}(T)$ also shows an activated behaviour at low temperatures, thus confirming the presence of a spin gap in AgVOAsO$_4$.
\end{abstract}
\maketitle

\section{Introduction}
Many quantum antiferromagnets have been recently reported, where strong quantum fluctuations destroy N$\acute{e}$el order and stabilize a singlet ground state. Such systems are characterized by a gap in their excitation spectrum. In gapped spin systems, the ground states are immune to perturbations such as inter-chain interactions, up to a threshold value, above which there is a transition to a long-range-ordered (LRO) state. External magnetic field has a somewhat similar effect. In low fields, the singlet state is retained, and the magnetization remains at zero. Higher fields close the gap in the excitation spectrum and also trigger long-range magnetic order that can be described as Bose-Einstein condensate (BEC) of triplons, magnetic excitations in a gapped quantum magnet. The BEC phenomenon has been actively studied in several model spin-$\frac12$ materials, such as TlCuCl$_3$ and BaCuSi$_2$O$_6$ representing three-dimensional (3D) and two-dimensional (2D) regimes of the BEC, respectively.\cite{Ruegg2003,Sebastian2006} Although the BEC transition does not occur in one dimension, quasi-1D materials with finite interchain couplings offer interesting effects in the applied magnetic field, including the Tomonaga-Luttinger liquid at higher temperatures and the BEC state at low temperatures.\cite{Klanjsek2008,Ruegg2008,Thielemann2009a,Thielemann2009b}

\begin{figure}[h]
\centering
\includegraphics[width=10cm]{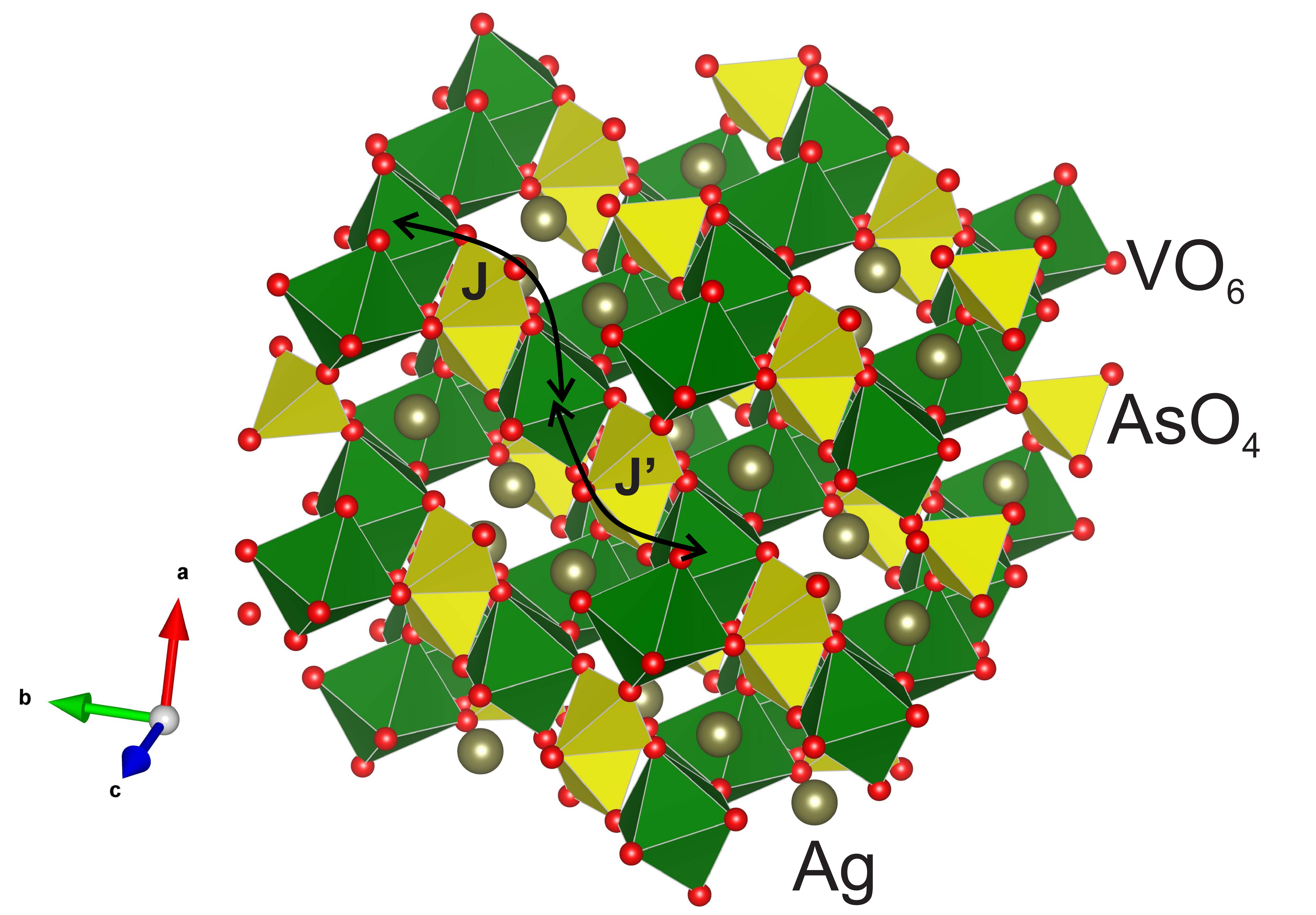}
\caption{\label{Fig1}The crystal structure of AgVOAsO$_{4}$. The VO$_{6}$ octahedra form alternating chains running along crossing directions and are inter-connected by AsO$_{4}$ tetrahedra. Thus, the As site is coupled to four V$^{4+}$ ions (two from each chain) belonging to two crossing magnetic chains, see Ref.~\onlinecite{Tsirlin2011} for further details.}
\end{figure}
AgVOAsO$_{4}$ crystallizes in a monoclinic unit cell (space group \textit{P}2$_{1}$/c) with lattice parameters $a \simeq 6.712$~{\AA}, $b \simeq 8.849$~{\AA}, $c \simeq 7.285$~{\AA}, and $\beta \simeq 115.28^{\circ}$. The crystal structure of AgVOAsO$_{4}$ shown in Fig.~\ref{Fig1} contains one vanadium and one arsenic site. The vanadium atoms form distorted VO$_{6}$ octahedra with V-O distances varying from 1.64 to 2.14~{\AA}. The arsenic atoms form nearly regular AsO$_{4}$ tetrahedra with As-O distances of 1.69-1.71~{\AA}. The chains are formed along the crystallographic $c$-axis via corner sharing of the VO$_{6}$ octahedra, whereas the AsO$_{4}$ tetrahedra connect the neighboring chains into a 3D structure.

Magnetic susceptibility $\chi(T)$, high-field magnetization, Electron Spin Resonance (ESR), and microscopic band structure calculations for AgVOAsO$_4$ have been reported in Ref.~\onlinecite{Tsirlin2011}. Though the $\chi(T)$ data show a broad maximum at around 20~K, a hall-mark of low-dimensionality, bulk magnetic susceptibility $\chi(T)$ at low temperatures is dominated by a Curie-like paramagnetic contributions of unknown origin. An alternating spin-$\frac{1}{2}$ chain model fit yields the exchange couplings $J/k_{\rm B} \simeq 41.8$~K and $J^{'}/k_{\rm B} \simeq 25.8$~K. The low-temperature magnetization curve corrected for the paramagnetic impurity contribution reveals that the intrinsic magnetization remains zero up to 10~T, which is the critical field ($H_{\rm c1}$) of the gap closing. The magnetization saturates at $H_{\rm c2} \simeq 48.5$~T. From the value of $H_{\rm c1}$, the spin gap was estimated to be $\Delta/k_{\rm B} \simeq 13$~K.\cite{Tsirlin2011}

The microscopic band structure calculations predict that alternating spin chains run along the [110] and [1$\overline{1}$0] directions, whereas interchain couplings are frustrated. Surprisingly, the structural chains were found to be not representing the direction of the leading exchange interactions, because no overlap between the $d_{\rm xy}$ orbital of V$^{4+}$ and $p$-orbital of the bridging oxygen atom is possible. Stronger superexchange interactions were found to run perpendicular to the structural chains via AsO$_{4}$ tetrahedra. Thus, each AsO$_{4}$ tetrahedra connects two VO$_{6}$ octahedra in one chain and two more VO$_{6}$ octahedra of another chain running perpendicular to the first one. Given the fact that both $H_{\rm c1}\simeq 10$\,T and $H_{\rm c2}\simeq 48.5$\,T are feasible for present-day high-field facilities, AgVOAsO$_4$ is an interesting model material for studying BEC in a quasi-1D system. However, its further investigation is hampered by the low-temperature Curie-like contribution that masks the putative spin gap. Such a contribution may arise from the magnetic anisotropy,\cite{Feyerherm2000} although it may also have a more trivial impurity origin.

In the following, we address the origin of this low-temperature contribution by $^{75}$As Nuclear Magnetic Resonance (NMR) measurements. We show that the intrinsic magnetic susceptibility of AgVOAsO$_4$ goes to zero at low temperatures, thus confirming the presence of a spin gap. NMR is a powerful microscopic tool to study the structural, static, and dynamic properties of frustrated spin systems. Since the $^{75}$As nucleus is inductively coupled to the magnetic spins, it is possible to extract information about the low-lying excitations of the V$^{4+}$ spins by measuring $^{75}$As NMR spectra, NMR shift, and spin-lattice relaxation time. In the polycrystalline sample, the presence of extrinsic impurities and defect spins normally hinders the analysis of $\chi(T)$ at low temperatures. In this context, the advantage of NMR is that the NMR line shift ($K$) is a direct measure of the \textit{intrinsic} spin susceptibility and is completely free from impurity contributions. For a random distribution of defect spins this paramagnetism broadens the NMR line but without contributing to the NMR shift. Thus, one can precisely estimate the magnetic parameters such as the exchange couplings and the spin gap from analyzing $K(T)$ instead of $\chi(T)$. Our analysis of $K(T)$ indeed confirms the alternating-spin-chain model with $J/k_{\rm B} \simeq 38.4$~K, $\alpha=J'/J\simeq 0.68$, and $\Delta/k_{\rm B} \simeq 15$~K. The exponential decrease of $K(T)$ at low temperatures provides a direct evidence of a spin gap and suggests that the upturn in $\chi(T)$ is extrinsic in nature. One can also estimate the spin gap by measuring the temperature-dependent spin-lattice relaxation time. In our case, it was calculated to be $\Delta/k_{\rm B} \simeq 15.9$~K.

\section{Experiments}
The synthesis of polycrystalline AgVOAsO$_{4}$ sample was done following the procedure reported in Ref.~\onlinecite{Tsirlin2011}.
The NMR experiments on $^{75}$As nucleus (nuclear spin $I = 3/2$ and gyromagnetic ratio $\gamma$/2$\pi$ = $7.291$~MHz/T) were carried out using pulsed NMR techniques at a radio frequency ($\nu$) of 49.5~MHz. The NMR spectra as a function of temperature $T$ were obtained by sweeping the magnetic field keeping the frequency ($\nu=\gamma/2\pi H$) fixed to 49.5~MHz. The NMR shift $K(T) = [H_{\rm ref}-H(T)]/H(T)$ was determined by measuring the resonance field $H(T)$ of the sample with respect to a standard GaAs sample (resonance field $H_{\rm ref}$). The spin-lattice relaxation rate ($1/T_{1}$) was measured using a conventional saturation pulse sequence.

\section{Results and Discussion}
\begin{figure}[h]
\centering
\includegraphics[width=7cm]{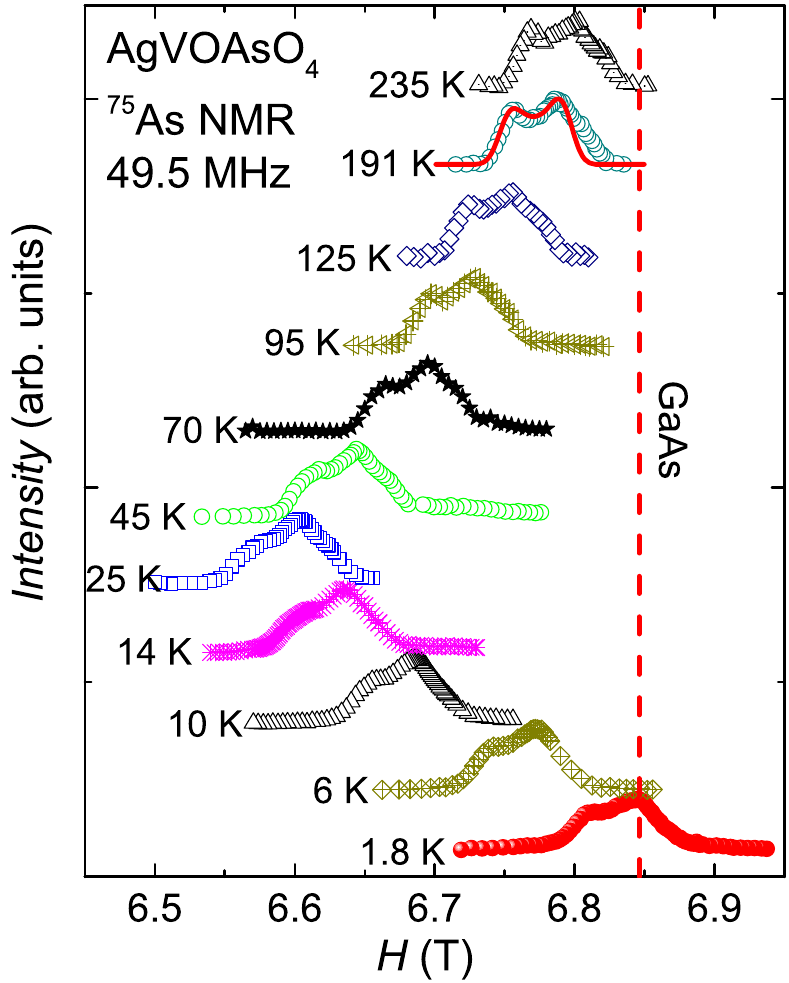}
\caption{\label{Fig2} Temperature-dependent field-sweep $^{75}$As NMR spectra of polycrystalline AgVOAsO$_{4}$ sample measured at 49.5 MHz. The vertical dashed line corresponds to the $^{75}$As resonance position of the reference GaAs sample. The solid line is the simulation of the spectrum at $T = 191$\,K with the parameters given in the text.}
\end{figure}
$^{75}$As is a quadrupole nucleus with the nuclear spin $I = 3/2$ and is located in a non-cubic environment in AgVOAsO$_{4}$. Therefore, the four-fold degeneracy of the nuclear spin $I=3/2$ is partially lifted by the interaction between the nuclear quadrupole moment $Q$ and the surrounding electric-field-gradient (EFG). In general, the nuclear spin Hamiltonian can be expressed as a summation of the Zeeman and nuclear quadrupole interaction terms as $H=-\gamma \hbar I \cdot H_{\rm eff} + \frac{h\nu_{Q}}{6}[3I_{z}^{2}-I(I+1)+\frac{1}{2}\eta(I_{+}^{2}+I_{-}^{2})]$, where $H_{\rm eff}$ is an effective field (a sum of the external field $H$ and hyperfine field $H_{\rm hf}$) at the $^{75}$As site, $h$ is the Planck's constant, $\nu_{Q}$ is nuclear quadrupole frequency defined as $\nu_{Q} = eQV_{ZZ}/2h$, where $V_{ZZ}$ is the EFG at the $^{75}$As site, and $\eta$ $(= |\nu_{Q}^{a}-\nu_{Q}^{b}|/\nu_{Q}^{c}$) is the asymmetry parameter of the EFG. When the quadrupole term is weak compared to the Zeeman term, it is enough to consider the effects up to first order in perturbation theory. In this case, two satellite peaks ($I_{z} = \pm 3/2 \longleftrightarrow \pm 1/2$) appear on either sides of the central line ($I_{z} = +1/2 \longleftrightarrow -1/2$), separated by $\nu_{Q}$, corresponding to three allowed transitions. The position of the satellite lines depends on the angle $\theta$ between the magnetic field direction and the direction of the largest EFG component $V_{\rm zz}$. When quadrupole effects are considered to second order (and for axial symmetry), the central line position (in the absence of anisotropy) also depends on $\theta$ and is given by the following equation:\cite{Slichter1992}
\begin{equation}
\label{quadrupole}
\nu_{(\pm\frac{1}{2})}^{(2)} = \nu_{0} + \frac{\nu_{\rm Q}^{2}}{32\nu_{0}}[I(I+1) - \frac{3}{4}](1 - \cos^{2}\theta)(9\cos^{2}\theta - 1),
\end{equation}
where $\nu_{0}$ the Larmor frequency. For a polycrystalline sample, the external field is oriented randomly and the spectrum is typically broad. The central line develops two peaks corresponding to $\theta \approx 41.8^{\circ}$ and $\theta = 90^{\circ}$.\cite{Grafe2008,Mukuda2008}

The typical $^{75}$As NMR spectra measured at different temperatures are depicted in Fig.~\ref{Fig2}. Our $^{75}$As NMR spectrum corresponds to the central transition and is well described in the whole measured temperature range by the second-order nuclear quadrupolar interaction. The double-horn spectrum at $T = 191$~K could be fitted reasonably well with the parameters $K_{x} = K_{y} \simeq 0.25$\%, $K_{z} \simeq 0.612$\%, $\nu_{\rm Q} \simeq 6.29$~MHz, $\eta = 0$, and linewidth 103.5~kHz. Furthermore, the NMR line shape remains almost unchanged with temperature except for the line broadening. For a simple paramagnet, the line width at half maximum is expected to increase in a Curie-Weiss manner with decreasing temperature. We indeed observed a systematic line broadening with lowering temperatures. The quadrupolar frequency was found to remain almost unaltered in the whole measured temperature range indicating no further structural deformation of the VO$_{6}$ octahedra and hence rules out the possibility of any lattice distortion as was observed in case of spin-Peierls transition in CuGeO$_3$.\cite{Hase1993}

\begin{figure}[h]
\centering
\includegraphics[width=7cm]{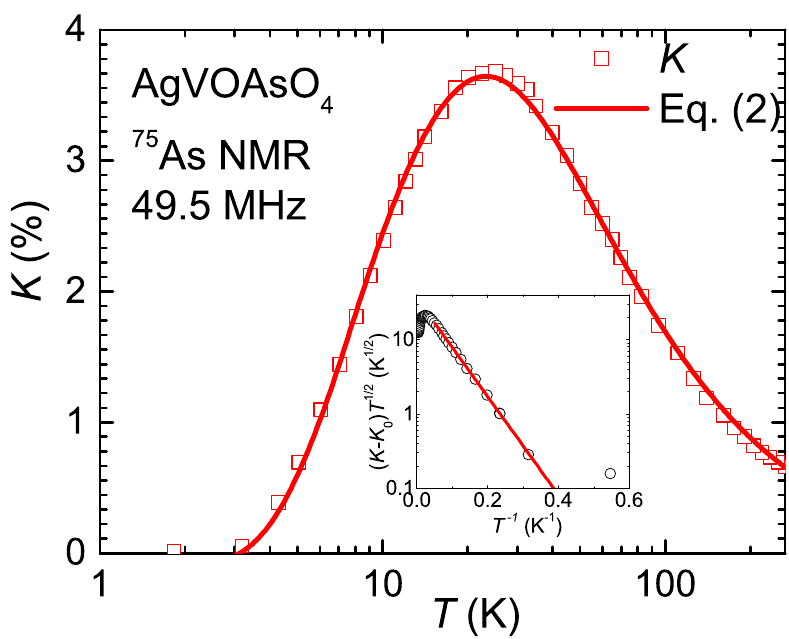}
\caption{\label{Fig3} Temperature-dependent $^{75}$As NMR shift $K$ as a function of temperature $T$. The solid red line is the fit of $K$ by Eq.~(\ref{shift}). Inset: $(K-K_{0})T^{1/2}$ vs $1/T$. The solid line indicates the activation law $(K-K_{0})T^{1/2} \propto exp(-\Delta/T)$ with $\Delta \simeq 15$~K. The least-square fit of the data was done below $T \leq 12$~K.}
\end{figure}
The line position is found to be shifting with temperature. The temperature-dependent NMR shift $K(T)$ was extracted by taking the field corresponding to the right-hand side of the most intense peak. It is presented in Fig.~\ref{Fig3}. It passes through a broad maximum at around 25 K, similar to the reported $\chi(T)$ data, indicative of the low-dimensional short-range ordering.\cite{Tsirlin2011} With decreasing temperatures, $K(T)$ is found to decrease rapidly and reach a constant value below 3~K. This sharp decrease of $K$ is a clear signature of the reduction of the V$^{4+}$ spin susceptibility within the chain and indicative of the spin gap between the singlet ($S=0$) ground state and triplet ($S=1$) excited states.

Since the NMR shift $K(T)$ is a direct measure of the spin susceptibility $\chi_{\rm spin}(T)$, one can write
\begin{equation}
\label{shift}
K(T) = K_{0} + \frac{A_{\rm hf}}{N_{\rm A}}\chi_{\rm spin}(T),
\end{equation}
where $K_{0}$ is the temperature-independent chemical shift and $A_{\rm hf}$ is the total hyperfine coupling constant between $^{75}$As nuclei and V$^{4+}$ electronic spins. $A_{\rm hf}$ includes contributions from transfered hyperfine coupling and the nuclear dipolar coupling, both of which are temperature-independent. The nuclear dipolar coupling is usually very small compared to the transfered hyperfine coupling and hence neglected. One can precisely estimate the exchange couplings by fitting $K(T)$ data with Eq.~(\ref{shift}). Here, $\chi_{\rm spin}$ is the expression for the spin susceptibility of the spin-$1/2$ Heisenberg alternating spin chain model given by Johnston \textit{et al.}\cite{Johnston2000}, valid over the whole temperature range of our experiment and also in the entire range of $0 \leq \alpha \leq 1$. Our $K(T)$ data were fitted by Eq.~(\ref{shift}) over the whole temperature range. In order to minimize the number of fitting parameters during the fitting procedure, the value of $g$ was fixed to $g \simeq 1.9$ obtained from the ESR experiments,\cite{Tsirlin2011} whereas $K_{0}$, $A_{\rm hf}$, $J_{1}/k_{\rm B}$, and $\alpha$ were varied. This value of $g$ is typical for V$^{4+}$ compounds.\cite{Ivanshin2003} As shown in Fig.~\ref{Fig3}, it fits well to our experimental data, especially in the broad maximum regime The fit yields $K_{0} \simeq -0.097$~\%, $A_{\rm hf} \simeq 3.5$~T/$\mu_{\rm B}$, the leading exchange coupling $J/k_{\rm B} \simeq 38.4$~K, and the alternation ratio $\alpha \simeq 0.68$, respectively. These values of $J/k_{\rm B}$ and $\alpha$ are in good agreement with those derived from the $\chi(T)$ analysis: $J\simeq 40$\,K and $\alpha=0.65$.\cite{Tsirlin2011}

As reported in Ref.~\onlinecite{Tsirlin2011}, the $\chi(T)$ data show a Curie-like upturn at low temperatures. In contrast, the exponential decrease observed in $K(T)$ clearly proves that the low-temperature upturn in $\chi(T)$ is extrinsic in nature. The Curie tail in $\chi(T)$ may be due to defects that break spin chains. If such defects are present in the sample, they will result in free spins and hence in an additional paramagnetic contribution. Also in powder samples, there are finite chains and the spins at the end of the chains produce staggered magnetization whose magnitude varies from one lattice site to another. This also contributes to the extrinsic paramagnetism.

The value of the spin gap $\Delta/k_{\rm B}$ can be estimated by analyzing $K$ in the temperature range $T \ll \Delta/k_{\rm B}$ based on the susceptibility of a gapped one-dimensional spin system.\cite{Sachdev1997,Damle1998} If the magnon dispersion along the chain is approximated by the quadratic form $\epsilon(k) \simeq \Delta + c^{2}k^{2}/2\Delta$ near the bottom of the dispersion, the temperature dependence of the susceptibility $\chi$ in the low-temperature limit $T \ll \Delta/k_{\rm B}$ is expressed as
\begin{equation}
\label{gap}
\chi = \sqrt{\frac{2\Delta}{\pi c^{2}T}}e^{-\Delta/T},
\end{equation}
where $c$ is the spin velocity. We fitted the $K(T)$ data below 12~K to the form $K=K_{0}+ b T^{-1/2}\exp(-\Delta/T)$. The obtained values are $K_{0} \simeq -0.106$~\%, $b \simeq 36.1$~K$^{1/2}$, and $\Delta/k_{\rm B} \simeq 15$~K. The results of the fit are shown in the inset of Fig.~\ref{shift} where we have plotted $(K-K_{0})T^{1/2}$ vs $1/T$. This value of $\Delta/k_{\rm B}$ is close to the that reported before ($\Delta/k_{\rm B} \simeq 13$~K).\cite{Tsirlin2011}

\begin{figure}[h]
\centering
\includegraphics[width=7cm]{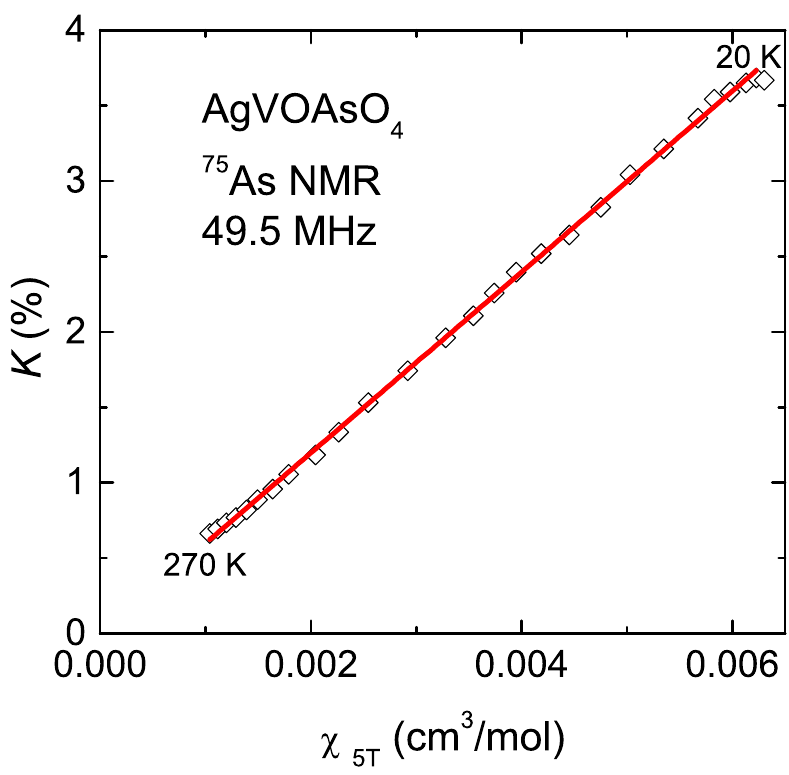}
\caption{\label{Fig4} The $^{75}$As NMR shift $K$ vs magnetic susceptibility $\chi$ measured at 5~T with temperature as an implicit parameter. The solid line is a linear fit.}
\end{figure}
According to Eq.~(\ref{shift}), one can also calculate $A_{\rm hf}$ by taking the slope of the linear $K$ vs $\chi$ plot. In Fig.~\ref{Fig4} we have plotted $K$ vs $\chi$ with $T$ as the implicit parameter. The $\chi(T)$ data used in Fig.~\ref{Fig4} were measured at 5~T, which is close to the field at which our NMR experiments were performed. Down to 20~K, it is a nearly straight line, while below 20~K, a deviation was observed which is because of the extrinsic paramagnetic contribution present in the $\chi(T)$ data. Our data in the temperature range 20~K to 270~K were fitted well to a linear function and the slope of the fit yields $A_{\rm hf}\simeq 3.3$~T/$\mu_{\rm B}$. This value of $A_{\rm hf}$ is in reasonable agreement with the value obtained from the $K$ vs $T$ analysis. However, it is an order of magnitude larger than the one observed for $^{31}$P in one-dimensional spin-$\frac12$ chain compounds like (Sr,Ba)Cu(PO$_4)_2$ and K$_2$CuP$_2$O$_7$ with similar interaction pathways.\cite{Nath2005,Nath2008b} Such a large hyperfine coupling suggests a strong overlap between the $p$-orbitals of As$^{5+}$ and $d$-orbitals of V$^{4+}$ ions via the $2p$ orbitals of O$^{2-}$. This also explains why the superexchange interaction between V$^{4+}$ ions is stronger via the V--O--As--O--V pathway than via the shorter V--O--V path.

\begin{figure}[h]
\centering
\includegraphics[width=7cm]{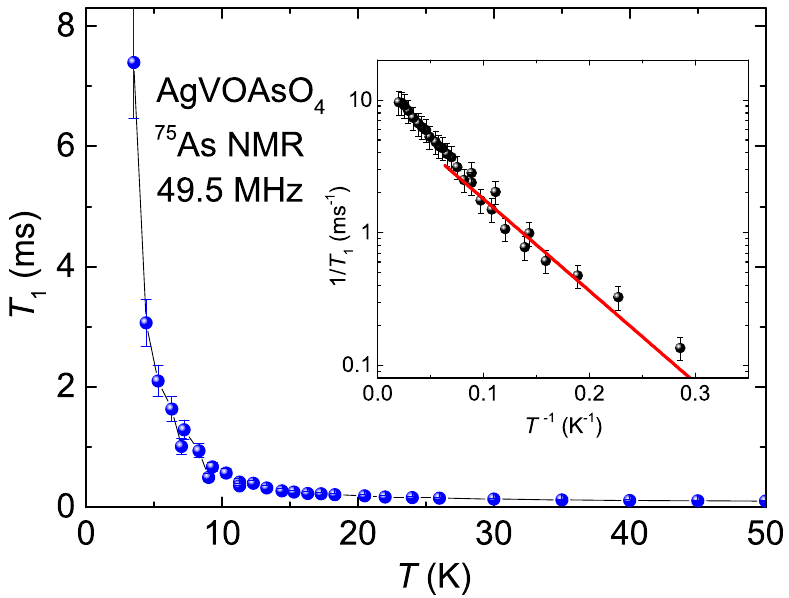}
\caption{\label{Fig5} $T_{1}$ as a function of temperature. Inset: $1/T_{1}$ is plotted against $1/T$. The solid line is the exponential fit in the low-temperature region.}
\end{figure}
One can also estimate the spin gap by analyzing the temperature-dependent spin-lattice relaxation rate $1/T_{1}$. The $^{75}$As spin-lattice relaxation time $T_{1}$ was measured at the field corresponding to the highest peak position. The recovery of the longitudinal magnetization at different temperatures after a group of saturation pulses was fitted by the following double-exponential function\cite{Gordon1978}
$1-\frac{M(t)}{M(\infty)} = 0.1\exp{(-t/T_{1})} + 0.9\exp{(-6t/T_{1})}$,
as expected for the central line of the spectrum of the nuclear spin $I = 3/2$ of the $^{75}$As nucleus.
Here $M(t)$ and $M(\infty)$ are the nuclear magnetizations at a time $t$ and at $t \rightarrow \infty$ (equilibrium), respectively after the saturation pulses. The $T_1$ estimated from the double exponential fit is plotted in Fig.~\ref{Fig5} as a function of $T$. At high temperatures, $T_{1}$ is approaching temperature-independent behaviour due to random fluctuations of paramagnetic moments.\cite{Moriya1956} It starts to increase slowly for $T \lesssim 30$~K which corresponds to the energy scale of the dominant exchange coupling $J_{1}/k_{\rm B}$.
Below about 15~K, $T_{1}$ was found to be increasing rapidly with temperature which is reminiscent of a spin-gap behaviour.

In the inset of Fig.~\ref{Fig5}, $1/T_{1}$ is plotted against $1/T$ and the $y$-axis is shown in log-scale in order to highlight the activated behaviour at low temperatures. We fitted the data between 3.5~K and 10.3~K by an activated form $1/T_{1} \propto \exp(-\Delta/k_{\rm B}T)$, which yields $\Delta/k_{\rm B} \simeq 15.9$~K. However, the absolute value of the gap determined from $1/T_{1}$ is slightly larger than the one obtained from the $K(T)$ analysis. Similar type of deviations were also observed in other gapped quantum spin systems such as Y$_{2}$BaNiO$_{5}$,\cite{Shimizu1995} CaV$_{2}$O$_{5}$,\cite{Iwase1996} SrCu$_{2}$O$_{3}$,\cite{Azuma1994} AgVP$_{2}$S$_{6}$,\cite{Takigawa1996} BaCu$_{2}$V$_{2}$O$_{8}$,\cite{Ghoshray2005} (VO)$_{2}$P$_{2}$O$_{7}$\cite{Kikuchi1999} etc.


\section{Conclusion}
Our $K(T)$ analysis unambiguously establishes that AgVOAsO$_{4}$ is an alternating spin-$\frac{1}{2}$ chain compound with $\alpha = J'/J \simeq 0.68$ and a spin gap $\Delta/k_{\rm B}\simeq 15$\,K. The exponential decrease of $K(T)$ at low temperatures directly confirms the existence of the spin gap. This also implies that the Curie like upturn observed in $\chi(T)$ was extrinsic in nature which is possibly arising from the defects present in the polycrystalline sample. The magnitudes of individual exchange couplings are consistent with our previous assessment based on the $\chi(T)$ and high-field magnetization measurements. The value of spin gap was calculated to be $\Delta/k_{\rm B} \simeq 15$~K and 15.9~K from the analysis of $K(T)$ and $1/T_1(T)$, respectively. From the fit of $^{75}$As NMR spectra, the quadrupole frequency was found to be $\nu_{\rm Q} \simeq 6.29$~MHz. The spectral shape remains almost intact over the whole measured temperature range, thus ruling out the possibility of any lattice distortion.

\acknowledgments
RN was funded by MPG-DST (Max Planck Gesellschaft, Germany and Department of Science and Technology, India) fellowship. AT was partly supported by the Federal Ministry for Education and Research through the Sofja Kovalevksya Award of Alexander von Humboldt Foundation.

\end{document}